\documentclass[12pt]{JHEP3}
\font\Nu=cmsy10 scaled 900
\def\Slash{\slash \kern-8pt}
\def\CP{{\bf CP}}

\def\be{{\begin{equation}}}
\def\ee{{\end{equation}}}
\def\longbar#1{\setbox1=\hbox{$#1$}
\setbox2=\vbox{\hrule width 0.8\wd1}
\raise0.5\ht1\hbox{${\lower\dp1\box2}\atop\box1$}}  
\def\mediumbar#1{\setbox1=\hbox{$#1$}
\setbox2=\vbox{\hrule width 0.6\wd1}
\raise0.5\ht1\hbox{${\lower\dp1\box2}\atop\box1$}}  

\title{The Standard Model Fermion Spectrum From Complex Projective Spaces}

\author{Brian P. Dolan and C. Nash\\
Dept. of Mathematical Physics, NUI, Maynooth, Ireland\\
{\rm and}\\
School of Theoretical Physics, Dublin Institute for Advanced Studies, 
10~Burlington Rd., Dublin 8, Ireland\\
{\tt bdolan@thphys.may.ie, cnash@thphys.may.ie}\\ }

\maketitle

\begin{abstract}{It is shown that the quarks and leptons 
of the standard model, including a right-handed neutrino, 
can be obtained by gauging the holonomy groups of complex
projective spaces of complex dimensions two and three.
The spectrum emerges as chiral zero modes of the Dirac operator coupled to 
gauge fields and the demonstration involves an index theorem analysis 
on a general complex projective space in the presence of topologically
non-trivial $SU(n)\times U(1)$ gauge fields.
The construction may have applications in type IIA string theory
and non-commutative geometry.}

\preprint{\tt DIAS-STP-02-07 \\}

\end{abstract}

\begin{document}

\section{Introduction}

It has been a long standing goal of theoretical particle physics to unify
space-time symmetries with the internal $SU(3)\times SU(2)\times U(1)$
gauge symmetry of the standard model.  Most current research in this
direction relies on string theory in higher dimensions, 
hoping to derive grand unified theories
as low energy limits of the full theory, and a crucial aspect of this programme
is the r\^ole of compact internal spaces.  Compact internal spaces were
first introduced into physics in Kaluza-Klein models where a coset space
$G/H$ with isometry group $G$ and holonomy group $H$ can give rise to a 
gauge group $G$ in 4-dimensional space-time.  A pure Kaluza-Klein approach was 
largely abandoned
in the 80s due in part to the realisation that it was difficult, if not
impossible, to obtain chiral Fermions this way \cite{ShelterIslandII}.
In this paper we take a different approach to internal coset spaces, focusing
on the holonomy group $H$ rather than $G$.  We shall show that a single generation
of the standard model spectrum, including a right handed neutrino, can arise
from the complex projective spaces $\CP^2$ and $\CP^3$.

The approach adopted here has the attractive feature that it is somewhat
more in tune with the spirit of general relativity than standard Kaluza-Klein theory.
In the standard approach one assumes that the coset space has a specific metric 
which has isometry group $G$, which is somewhat contrary to the philosophy
general relativity where a particular metric is merely one solution of
Einstein's field equations and there may be many others with smaller isometries,
indeed a generic solution has none.  In the construction used here it is the holonomy
group that is important and for a complex manifold with real dimension $2n$
this group is generically $U(n)$.  For 4-dimensional space time equipped
with a Lorentzian metric
the holonomy group is generically $SO(1,3)$ 
and general relativity can be viewed as a gauge theory of tangent space rotations
using this group.
Incorporating spinors into general relativity
requires gauging the double cover of $SO(1,3)$, namely
$Spin(1,3)\cong Sl(2,{\bf C})$. If a compact complex internal space with real
dimension $2n$ is added
the extra holonomy group is generically $U(n)$ and so it seems quite natural
to gauge $SU(n)\times U(1)$.  Taking two internal spaces with $n=2$ and $n=3$
would give $SU(3)\times SU(2)\times U(1)\times U(1)$ which is not quite what is
required, but is close.

To see how the standard model spectrum might arise in this way consider first a 
Dirac spinor on a general compact complex manifold of real dimension 
4. This has 4 components and decomposes as 
\begin{equation}
({\bf 2},{\bf 1}) + ({\bf 1},{\bf 2})\longrightarrow {\bf 2}_0 + ({\bf 1}_1 +{\bf 1}_{-1})
\end{equation}
under $Spin(4)\cong SU(2)\times SU(2)\rightarrow SU(2)\times U(1)$.
Shifting the $U(1)$ charge by $-1$ and rescaling the result by $1/2$ gives
\begin{equation}
{\bf 2}_{-1/2} + ({\bf 1}_0 +{\bf 1}_{-1}).
\label{eq:2-1-1}
\end{equation}
On the other hand a compact complex manifold of real dimension 
6 has 8 component Dirac spinors which decompose as 
\begin{equation}
{\bf 4} + \overline{\bf 4}\longrightarrow 
({\bf 3}_1 + {\bf 1}_{-3}) + (\overline{\bf 3}_{-1}+{\bf 1}_{3})
\end{equation}
under $Spin(6)\cong SU(4)\rightarrow SU(3)\times U(1)$.
Taking a single chiral spinor in the ${\bf 4}$ of $SU(4)$, shifting the $U(1)$ charge by $3$
and rescaling the result by $1/6$, gives
\begin{equation}
({\bf 3}_{2/3} + {\bf 1}_0).
\label{eq:3-1}
\end{equation}
A single generation of the standard model spectrum,
including a right-handed neutrino, can be obtained by tensoring (\ref{eq:2-1-1})
with (\ref{eq:3-1}) and simply adding the $U(1)$ charges.
These group theory decompositions are valid on any complex manifolds of real
dimension 4 and 6 respectively and the shift in the $U(1)$ charges may be achieved
by coupling the Fermions to $U(1)$ gauge fields with appropriate topological (monopole)
charges. Mathematically this requires tensoring the spin bundle with an
appropriate line bundle, which may or may
not be possible on any given manifold depending on whether or not appropriate
line bundles exist.  It is shown in section \ref{sec:CPN} that for $\CP^2$ and $\CP^3$ the required
line bundles do indeed exist 
(indeed for $\CP^2$ spinors cannot even be defined unless such a line
bundle is introduced).
However obtaining the standard
model spectrum requires introducing some natural higher rank bundles as well.
Furthermore, an index theorem analysis 
shows that the representations (\ref{eq:2-1-1}) and (\ref{eq:3-1})
can be realised as zero modes of the Dirac operator with precisely the correct handedness for
one generation of the standard model, including a right-handed neutrino.
This is a topological statement, it is completely independent of the choice
of metric on these spaces.

Of course there is the question of generations---the construction
presented here only produces a single generation of the standard model.
This point is elaborated on in the concluding section \ref{sec:conc},
along with some suggestions of how the construction might fit into
viable models such as type IIA superstring theory and non-commutative geometry. 
Details of the index theorem on $\CP^n$, required for the analysis in the
text, are contained in an appendix, where closed form expressions for the
index of the Dirac operator coupled to topologically non-trivial $SU(n)\times U(1)$
gauge fields are derived.

\section{The Particle Spectrum From $\CP^n$}
\label{sec:CPN}

In this section it will be argued that the spectrum of a single generation of the
standard model can be obtained from $\CP^2\times \CP^3$ by  
gauging the holonomy group, $U(2)\times U(3)$,
with the two $U(1)$ factors identified in one particular combination.
This generalises the case of $\CP^2$ which gives the electroweak sector
\cite{FuzzySpinc}.

The analysis is based on the index theorem for the
Dirac operator on 
\begin{equation}
\CP^n\cong {SU(n+1)\over U(n)}
\end{equation}
 derived in the appendix.  
We are interested in zero modes of the Dirac operator, for Fermions
which are either singlets or transform under the fundamental ${\bf n}$
representation of $SU(n)$, when topologically non-trivial gauge
fields are present ($\CP^n$ analogues
of the monopole field on $S^2$ and the $SU(2)$ BPST instanton on $S^4$).
The gauge group is taken to be the holonomy group $U(n)$ of $\CP^n$.
The index is the difference between the number
of positive chirality zero modes of the Dirac operator and the number of negative chirality
zero modes and, for convenience,
the important formulae from the appendix are reproduced here. For a singlet with $U(1)$
charge $Y_{(n)}=q$, where $q$ is an integer, the index is
\begin{equation}
 \nu_q={1\over n!}(q+1)\cdots(q+n),\,q\in{\bf Z}
\end{equation}
corresponding to a line bundle over $\CP^n$ with gauge group $U(1)$.
A Fermion in the fundamental representation of $SU(n)$, with a $U(1)$ charge
$Y_{(n)}=q+{1\over n}$, has index
\begin{equation}
\nu_{q,{\bf n}}={(q+1)\cdots\bigl(q+(n-1)\bigr)\bigl(q+n+1\bigr)\over (n-1)!}.
\end{equation}
We are free to rescale the $U(1)$ charges differently for different values of $n$,
but for a given $n$ the ratio of the singlet charge $q$ to 
the fundamental charge $q+{1\over n}$ is fixed.

The simplest case is $\CP^1\cong S^2$, where the holonomy group is $U(1)$.
Introducing $U(1)$ gauge fields and Fermions with charge $q$ 
there are $\nu_q=q+1$ zero modes of the Dirac operator.  
The case $q=-1$ gives no unpaired zero modes---generically there are none at all
but even when zero modes exist left and right-handed particles with charge $-1$
occur in pairs, like an electron in QED.   
For $q\ne -1$ the theory is necessary chiral.
For example choosing the convention that positive $\nu$ corresponds
to right-handed spinors in $4$-dimensions, 
$q=-2$ would be interpreted as a left-handed particle with hypercharge $-2$.
However there
is no place in this construction for a right-handed particle with the same charge as this
would require $q=-2$ and $\nu>0$.  This is necessarily a chiral theory 
for $q\ne -1$,
and would generate a lethal gauge anomaly in $4$-dimensions.

To introduce quarks we turn to $\CP^3$---since
the holonomy group of $\CP^3$ is $U(3)$ quarks can be introduced as
triplets.  For example an $SU(3)$ triplet 
with $q=-3$ gives 
$\nu_{-3,{\bf 3}}=1$ and has $U(1)$ charge $Y_{(3)}=-8/3$. 
Since the overall normalisation of the charge is at our disposal, this
could represent either: 
\begin{equation}
 \hbox{a right handed d-quark (rescale charge by $1/8$),} \qquad d_{\bf R}={\bf 3}_{-1/3},
 \label{eq:RHd}
\end{equation}
or 
\begin{equation}
 \hbox{a right-handed u-quark (rescale charge by $-1/4$),} \qquad u_{\bf R}={\bf 3}_{2/3}.
 \label{eq:RHu}
\end{equation}
Either possibility forces us to interpret
positive chirality as corresponding to right-handed Fermions in $4$-dimensions
in order to match the standard model spectrum.
It is instructive to examine the complex conjugate representations in order
to understand the CPT conjugate of the spectra.
In the appendix it is shown that under complex conjugation of the bundles over $\CP^n$
the $U(1)$ charges transform as 
$Y_{(n)}\rightarrow \overline Y_{(n)}=-Y_{(n)}-(n+1)$ and the index
as $\nu\rightarrow\overline\nu=(-1)^n\nu$.  
So for the examples (\ref{eq:RHd}) and (\ref{eq:RHu}) 
above, with just a single $SU(3)$ triplet on $\CP^3$
with $Y_{(3)}=-8/3$, $\overline Y_{(3)}=-4/3$ and
complex conjugation flips the chirality so it maps
\begin{equation}
 d_{\bf R}={\bf 3}_{-1/3}  \longrightarrow 
 \overline {\bf 3}_{-1/6}
\end{equation}
and 
\begin{equation}
 u_{\bf R}={\bf 3}_{2/3} \longrightarrow 
 \overline {\bf 3}_{1/3}=(\overline d)_{\bf L}.
\end{equation}
If we start with a right-handed $u$-quark complex conjugation 
forces us to introduce the left-handed anti-$d$ while
starting with a right-handed $d$-quark complex conjugation forces the introduction
of a state that has no place in the standard model.  This nicely illustrates the kind
of constraints that CPT can place on the possible choices.

For weak interactions we bring in $\CP^2$ with holonomy group $U(2)$. 
An index theorem analysis then allows us to 
obtain a single generation of the electroweak sector, including a right-handed
neutrino.  This is achieved by taking the following  three representations:
\par\vskip0.15\baselineskip
\begin{itemize}
\item{} An $SU(2)$ singlet with $q=0$
giving zero charge and  index $\nu_0=+1$; 
\item{} A second singlet with $q=-3$ giving charge $Y_{(2)}=-3$ and index $\nu_{-3}=+1$; 
\item{} An $SU(2)$ doublet with $q=-2$ giving charge $Y_{(2)}=-3/2$ and $\nu_{-2,{\bf 2}}=-1$.
\end{itemize}
\par\vskip0.15\baselineskip
Interpreting positive $\nu$ as giving right-handed spinors, and rescaling the charge
by $1/3$, this results in 
a single generation of particles of the electroweak sector of the
standard model, including a right-handed neutrino:
\begin{equation} 
 {\bf 1}_0=({\hbox{\Nu V \,}})_{\bf R} 
 \qquad {\bf 1}_{-1}=e_{\bf R} \qquad 
 {\bf 2}_{-1/2}=\left(\matrix{\hbox{\Nu V \,}_{\bf L}\cr
 e_{\bf L}}\right)
 \label{eq:EWL}
\end{equation}
(the normalisation is such that the electric charge is $Q=I_3 + Y$).
Now for $\CP^2$ complex conjugation preserves the chirality and sends $Y_{(2)}\rightarrow
\overline Y_{(2)}=-Y_{(2)}-3$, so complex conjugation maps (\ref{eq:EWL}) to
\begin{equation} 
 {\bf 1}_{-1}=e_{\bf R} \qquad 
 {\bf 1}_0=({\hbox{\Nu V \,}})_{\bf R} 
 \qquad 
 \overline {\bf 2}_{-1/2}=\left(\matrix{e_{\bf L}\cr
 \hbox{\Nu V \,}_{\bf L}}\right)
\label{eq:EWR}
\end{equation}
(for the $\overline{\bf 2}$ the electric charge is $Q=-I_3 + Y$ and of course the
$\overline{\bf 2}$ is equivalent to the ${\bf 2}$ under rotation
by the Pauli matrix $i\sigma_2$).  The curious conclusion is that,
contrary to what one might na{\"\i}vely expect, complex conjugation
does not change the sign of the hypercharge but instead interchanges
the electron and neutrino, leaving the electroweak
multiplet in (\ref{eq:EWL}) invariant.

%On the other hand, interpreting positive $\nu$ as giving left-handed spinors
%in agreement with the quark construction on $\CP^3$ above, and rescaling the charge
%$Y_{(2)}$ by $-1/3$ rather than $1/3$ these three representations on $\CP^2$ give
%a single generation of anti-particles of the electroweak sector of the
%standard model, including a left-handed anti-neutrino:
%\begin{equation} {\bf 1}_0=(\overline{\hbox{\Nu V \,}})_{\bf L} 
%\qquad {\bf 1}_{1}=(\overline{e})_{\bf L} \qquad 
%{\bf 2}_{1/2}=\left(\matrix{(\overline{e})_{\bf R}\cr
%(\overline{\hbox{\Nu V \,}})_{\bf R}}\right).
%\label{eq:EW}\end{equation}
%Again complex conjugation maps this multiplet onto itself, but interchanges
%the positron and the anti-neutrino.

One complete generation of the quark sector of the standard model can be obtained
by combining (\ref{eq:EWL}) with the $SU(3)$ triplet in (\ref{eq:RHu}),
provided the total hypercharge is defined  as a particular linear combination of the
$\CP^2$ and the $\CP^3$ charges, $Y=-{1\over 4}Y_{(3)}+{1\over 3}Y_{(2)}$.
Taking the total chirality to be the product of the two individual chiralities, and interpreting
positive chirality as right-handed, gives the particle spectrum of the strong sector
of the standard model:
 
\begin{equation} 
 ({\bf 3},{\bf 1})_{2/3}=u_{\bf R} \qquad 
 ({\bf 3},{\bf 1})_{-1/3}=d_{\bf R} \qquad 
 ({\bf 3},{\bf 2})_{1/6}=\left(\matrix{u_{\bf L}\cr 
 d_{\bf L}}\right).
 \label{eq:SMQCD}
\end{equation}

The electroweak sector can be included by combining an $SU(3)$ singlet
on $\CP^3$, with zero charge and $\nu_0=1$, with (\ref{eq:EWR}):
\begin{equation} 
 ({\bf 1},{\bf 1})_0=\hbox{\Nu V \,}_{\bf R} 
 \qquad ({\bf 1},{\bf 1})_{-1}=e_{\bf R} \qquad 
 ({\bf 1},{\bf 2})_{-1/2}=\left(\matrix{
 \hbox{\Nu V \,}_{\bf L}\cr 
 e_{\bf L}\cr }\right).
 \label{eq:SMEW}
\end{equation}
Equations (\ref{eq:SMQCD}) and (\ref{eq:SMEW}) constitute a 
single generation of the standard model.

Now observe that 
the combination $Y={1\over 4}Y_{(3)}-{1\over 3}Y_{(2)}$ {\it does} simply change
sign under complex conjugation, $Y \rightarrow \overline Y =-Y$.  In addition the $\CP^3$
chirality changes while that of $\CP^2$ does not so the overall chirality,
which is the product of the two, flips.  The net result is that complex conjugation
of (\ref{eq:SMQCD}) and (\ref{eq:SMEW}) does indeed reproduce the anti-particles:

\begin{eqnarray}
 ({\bf 3},{\bf 1})_{2/3}&=u_{\bf R}, \quad  
 ({\bf 3},{\bf 1})_{-1/3}&=d_{\bf R}, \quad 
 ({\bf 3},{\bf 2})_{1/6}=
  \left(\matrix{u_{\bf L}\cr
  d_{\bf L}\cr}\right)\\
 \longmapsto\raise -7pt \hbox{\kern -19pt c.c.}\qquad
 (\overline{\bf 3},{\bf 1})_{-2/3}&=(\overline u)_{\bf L}, \quad  
 (\overline{\bf 3},{\bf 1})_{1/3}&=(\overline d)_{\bf L}, \quad
  (\overline{\bf 3},\overline{\bf 2})_{-1/6}=
  \left(\matrix{(\overline u)_{\bf R}\cr 
  (\overline d)_{\bf R}}\right)\nonumber
\end{eqnarray}
and
\begin{eqnarray} 
({\bf 1},{\bf 1})_0&=\hbox{\Nu V \,}_{\bf R}, 
 \qquad ({\bf 1},{\bf 1})_{-1}&=e_{\bf R}, \quad 
 ({\bf 1},{\bf 2})_{-1/2}=\left(\matrix{\hbox{\Nu V \,}_{\bf L}\cr
  e_{\bf L}\cr }\right)\\
\longmapsto\raise -7pt \hbox{\kern -19pt c.c.}\qquad
 ({\bf 1},{\bf 1})_0&=(\overline{\hbox{\Nu V \,}})_{\bf L}, \quad 
 ({\bf 1},{\bf 1})_{1}&=(\overline{e})_{\bf L}, \quad 
  ({\bf 1},\overline {\bf 2})_{1/2}=
  \left(\matrix{(\overline{\hbox{\Nu V \,}})_{\bf R}\cr
  (\overline{e})_{\bf R}\cr}\right).\nonumber
\end{eqnarray}

It is interesting that the construction presented here necessarily requires
the introduction of a right-handed neutrino, as recent experimental
evidence for neutrino oscillations requires just such a state for
the simplest explanation of the results \cite{SK}, \cite{SNO}.

In summary a single complete generation of the standard model, (\ref{eq:SMQCD}) and (\ref{eq:SMEW}),
arises from a selection of five different bundles, three over $\CP^2$ and two over $\CP^3$:
\par\vskip0.15\baselineskip
\begin{itemize}
 \item{} Two $SU(2)$ singlets on $\CP^2$ with $q=0$ and $q=-3$ 
and an $SU(2)$ doublet with $q=-2$;

\item{} An $SU(3)$ singlet on $\CP^3$ with $q=0$ and an $SU(3)$ triplet with $q=-3$.
\end{itemize}
\par\vskip0.15\baselineskip

In terms of the $Spin^c$ structures described in the appendix
these correspond to the two bundles
\begin{eqnarray}
\hbox{on $\CP^2$}\qquad\qquad&&\wedge^{0,*}T\CP^2\otimes\left({\bf 1}
\oplus L_{(2)}^3\oplus \left(F_{(2)}\otimes L_{(2)}^2\right)\right)
\label{eq:redCP2}\\
\hbox{on $\CP^3$}\qquad\qquad&&\wedge^{0,*}T\CP^3\otimes \left({\bf 1}\oplus \left(F_{(3)}\otimes L_{(3)}^3\right)\right)\label{eq:redCP3}
\end{eqnarray}
where, on $\CP^2$,  $L_{(2)}$ is the generating line bundle  
and $F_{(2)}$ the rank 2 bundle
satisfying $L_{(2)}\oplus F_{(2)}=I^3$, while, on $\CP^3$,
$L_{(3)}$ is the generating line bundle and $F_{(3)}$ the rank 3 bundle
satisfying $L_{(3)}\oplus F_{(3)}=I^4$, with $I^3$ and $I^4$ denoting trivial
bundles of rank 3 and 4 respectively.

\section{Conclusions}
\label{sec:conc}
It has been shown that a single generation of the standard model Fermion
spectrum, including a right-handed neutrino, can be obtained from the
holonomy groups of $\CP^2$ and $\CP^3$ by tensoring with appropriate
bundles.  Physically this means introducing background $SU(3)\times SU(2)\times U(1)$
gauge fields which are topologically non-trivial, containing analogues of monopoles
and instantons.
A number of questions present themselves.

Firstly there is no obvious sign of three generations.
Of course one can obtain more generations by 
taking copies,
but there seems no compelling reason to take three such copies
and not some other number.
This may be related to the question of what possible r\^ole
the isometry group may play.  In the introduction a virtue was made of the
fact that the construction is not tied down to a specific metric on $\CP^n$,
but if one introduces the Fubini-Study metric one then has isometry group $SU(n+1)$.
On $\CP^2$ one has $SU(3)$ and, using this as a horizontal
generation group, the fundamental representation would give three generations.
But then it is not clear what the r\^ole of the $SU(4)$ from $\CP^3$ would be.
Alternatively it is possible to manufacture three copies by including $\CP^1$ with
$q=2$, giving $\nu_2=3$, and then simply ignoring the $U(1)$ charge
on $\CP^1$.  For the moment we have no compelling
suggestion as to how the generations might appear and we leave this as an
open question.

Secondly there is the question of the extra $U(1)$.  The holonomy
group of $\CP^2\times \CP^3$ is $U(3)\times U(2)\cong SU(3)\times SU(2)\times U(1) \times U(1)$
and we have taken a $U(1)$ which is only one particular combination of the two $U(1)$ 
factors, ignoring the other one.  In fact the standard model spectrum has true
group $S\bigl(U(3)\times U(2)\bigr)$, \cite{Lochlainn}, and the very fact that it is reproduced
here means that the group being used is $S\bigl(U(3)\times U(2)\bigr)$ rather than the full
holonomy group $U(3)\times U(2)$.  There may be a deeper reason for this,
but for the moment we confine ourselves to observing that it works empirically.

How might the construction fit into realistic models?
It would certainly seem too na\"\i ve to take $\CP^2\times \CP^3$
as an internal space---it has real dimension 10 which is too large for string
theory and simply adding it on to 4-dimensional space-time produces a 14-dimensional
space-time which would have anomalies.  It may be that one could realise $\CP^2$ and
$\CP^3$ as a brane within a brane in type IIA string theory, which
has anti-symmetric tensor fields of rank 2 and 4 in its R-R sector
as well as their hodge duals, though IIA string string theory is a
non-chiral theory so the Weyl Fermion on $\CP^3$ would have to
be put in by hand.  
Any such interpretation would necessarily
be rather different to the standard approach, as it would not involve grand
unified theories directly.  Alternatively a ``fuzzy'' Kaluza-Klein
approach may be of interest where the continuum manifolds of $\CP^2$ and $\CP^3$ are replaced
by non-commutative 
finite dimensional matrix approximations with a finite number of degrees of freedom
\cite{FuzzySpinc}.
It would then be more appropriate to think of multiple copies of $4$-dimensional
space-time rather than an internal continuous manifold, 
somewhat analogous to
Connes' non-commutative geometry approach to the standard model
with two copies of space-time \cite{ConnesLott}, \cite{Connes}.
Star-products on $\CP^n$ were studied in \cite{Bordemann}
and \cite{FuzzyCPN}, and the spectrum of the Dirac operator
was investigated in \cite{Grosse} and \cite{BalCP2}.
The smallest vector space that could be used for a non-trivial matrix
representation is $n+1$.  For $\CP^2$  we get 3, which relates to the
discussion of the generation problem above, while $\CP^2\times \CP^3$ would require $3\times4=12$
copies.

While these are all interesting and important problems we leave them open for
further work.

\appendix
\section{The Index Theorem For $\CP^n$}
For a general complex manifold $X$ of dimension $n$ the total Chern 
class is the 
sum of the individual Chern classes
\begin{equation}
c(X)=1+c_1(X)+ c_2(X)+\cdots  +c_n(X).
\end{equation}
In line with common usage we write $c_k(X)$ for $c_k(TX)$---the 
$k^{th}$ Chern class
of the tangent bundle.  In particular $c_n(X)$
is the top form and the Gauss--Bonnet theorem states that 
evaluating $c_n(X)$ on $X$ gives the Euler characteristic of $X$: i.e.
$c_n(X)[X]=\chi(X)$.
\par
For $\CP^n$ the Chern classes are all generated by a single $2$ dimensional class  $x$
\cite{Bott+Tu}
\begin{equation}
c(\CP^n)=(1+x)^{n+1}=1+(n+1)x +{n(n+1)\over 2}x^2 +\cdots +(n+1)x^n.
\end{equation}
Note that $x^{n+1}$  is a $(2n+2)$ dimensional class and so 
$x^{n+1}=0$---when $x$ is represented by a 2 form $\omega$, 
say, this corresponds to the fact that  $\omega^{n+1}=0$  
on a $2n$-dimensional manifold.
\par
The normalisation is such that
\begin{equation}
\int_{\CP^n} \omega^n=1
\end{equation}
and so, since $c_n(X)=(n+1)x^n\equiv (n+1)\omega^n$,  we have
\begin{equation}
\chi(\CP^n)=n+1,
\end{equation} 
which is the Euler characteristic of $\CP^n$ as required by the Gauss-Bonnet theorem.
The form $\omega$ is the curvature of a line bundle $\overline L$ 
whose complex conjugate $L$ we shall refer to as
the {\it generating line bundle} over $\CP^n$ 
with Chern class
\begin{equation}
c(L)=1-x.
\end{equation}

Another line bundle that will be important is the 
canonical line bundle $K$ which is the maximum exterior power 
of the cotangent bundle $T^*X$: i.e. 
\begin{equation}
K=\wedge^n T^*X.
\end{equation} 
$K$ has  Chern class given by
\begin{equation}
c(K)=1-(n+1)x.
\end{equation}
The minus sign appears because 
\begin{equation}
c_1(K)=c_1(T^*X)=-c_1(TX).
\end{equation}
We see that $K$ is a power of the generating line bundle $L$; in fact we have
\begin{equation}
K=L^{n+1}.
\end{equation}
\par
The existence of global spinors   is determined by the
second Stiefel--Whitney class $w_2$, which,  on a complex manifold $X$, 
can be obtained from $c_1$  by reducing mod $2$.  
One has
\begin{equation}
w_2(X)=c_1(X)\;{\rm mod}\,2\quad \hbox{($X$ complex)}.
\end{equation}
For $X=\CP^n$ we see that
\begin{eqnarray}
w_2(X)&=&(n+1)x\;{\rm mod}\,2,\quad X=\CP^n\\
      &=&\cases{0,& if $n$ is odd\cr
                \not=0,& if $n$ is even.\cr}
\end{eqnarray}
Hence $\CP^n$ admits globally defined spinors for odd $n$ but not for even $n$.
\par
When $n$ is even it is  still possible to define what is called a 
$Spin^c$ structure and this gives a more general kind  of spinor which 
comes accompanied by a line bundle and hence a possible $U(1)$ connection. 

Moreover, when, as is the case here,  $X$ is a complex manifold 
both spin structures and $Spin^c$ structures have a concrete form in terms of 
differential forms which goes as follows: consider the bundle 
$\wedge^{0,*}TX$ of all forms of type $(0,k)$---i.e. anti-holomorphic 
$k$ forms---so we have
\begin{equation}
\wedge^{0,*}TX={\textstyle\bigoplus\limits_k}\wedge^{0,k}\longbar{T^*X}.
\end{equation}
Then this is the $Spin^c$  bundle $S^c(X)$  
 i.e.
\begin{equation}
S^c(X)=\wedge^{0,*}TX.
\end{equation}
Hence, for a complex K\"ahler manifold, generalised spinors are  $(0,k)$ 
forms and are sections of $S^c(X)$; the Dirac operator is then the  
operator $\bar \partial+\bar \partial^*$.
Now if $X$ is not just a $Spin^c$ manifold but also a spin manifold 
then it has a bundle $S(X)$ of true spinors given by
\begin{eqnarray}
S(X)&=&S^c(X)\otimes K^{1/2}\\
\Rightarrow S^c(X)&=& S(X)\otimes K^{-1/2}.
\label{eq:SK}\end{eqnarray} 
Now the point is that the LHS of \ref{eq:SK} always exists 
but the two bundles $S(X)$  and $K^{-1/2}$ on the RHS only 
exist separately when $X$ is a spin manifold. One 
can still think of sections of $S^c(X)$ as being given 
`locally' by sections of 
$S(X)\otimes K^{-1/2}$---hence the interpretation  of 
Spin$^c$ structures as being (locally) 
a spinor with an attendant $U(1)$ gauge connection.
\par
Specialising to the case 
\begin{equation}
X=\CP^n
\end{equation}
we see that $S^c(\CP^n)=\wedge^{0,*}T\CP^n$ exists for all $n$ but 
$S(\CP^n)$ and $K^{-1/2}$ exist only for odd $n$.
\par
We can construct more general $U(1)$ bundles by adding more integral powers of $L$ giving 
\begin{eqnarray}
&&S^c(X)\otimes L^{-q}\\
&=&\wedge^{0,*}TX\otimes L^{-q},\quad X=\CP^n\\
&=&S(X)\otimes L^p,\quad\hbox{ where }p=-q-{n+1\over 2}
\end{eqnarray}
and we have used the fact that $K=L^{n+1}$; note too that $p$ is an integer for 
$n$ odd and a half-integer for $n$ even (the minus sign in the exponent $L^{-q}$
is for later convenience).

In physics language tensoring with powers of $L$ means introducing a $U(1)$ gauge field
and $q$ will be related to the $U(1)$ charge.   It is analogous to a monopole charge
which is evaluated by integrating the Chern character $ch(L^{-q})$ over a non-trivial
2-cycle (a $2$-sphere) in $\CP^n$.  Our conventions are such that
\begin{equation}
\int_{S^2}ch(L)=\int_{S^2}c_1(L)=-1,
\end{equation}
so
\begin{equation}
\int_{S^2}ch(L^{-q})=-q\int_{S^2}c_1(L)=q.
\end{equation}

For a given $p$ the net number of zero modes of the Dirac operator is given
by the index theorem as $\nu$ where 
\begin{equation}
\nu=\int_X ch(L^p)\hat A(X),\quad X=\CP^n.
\end{equation}
This equation can be written solely in terms of the K\"ahler $2$-form 
$\omega$: since the $\hat A$-genus and the Chern 
character for $\CP^n$ are given by
\begin{eqnarray}
\hat A(X)&=&\left({\omega\over 2\sinh(\omega/2)}\right)^{n+1}\qquad
\hbox{and}\qquad ch(L^p)=e^{-p\omega}\\
\Rightarrow ch(L^p)\hat A(X)&=&
e^{-p\omega}\left({\omega\over e^{\omega/2}- e^{-\omega/2}}\right)^{n+1},\quad X=\CP^n \\
&=&e^{q\omega}\left({\omega \over 1-e^{-\omega}}\right)^{n+1}\qquad \hbox{since}\quad p=-q-{n+1\over 2}.
\end{eqnarray}
Since   $\int_{\CP^n}\omega^n=1$ the index, which we shall denote by $\nu_q$, is the coefficient of 
 the $\omega^n$ term. This coefficient can be evaluated by
integration around a small contour enclosing the origin in the complex $z$-plane giving 

\begin{equation}
\nu_q={1\over 2\pi i}\oint {e^{qz} dz\over z^{n+1}} 
\left({z \over 1-e^{-z}}\right)^{n+1}.
\end{equation}
A change of variable to $t=1-e^{-z}$ yields the answer 
\begin{equation}
 \nu_q=\left(
    \matrix
       {q+n \cr 
       n \cr}
  \right)={1\over n!}(q+1)\cdots(q+n).
\label{eq:index}
\end{equation}
This formula works for both even and odd $n$ and for any integral $q$, either positive or negative.
The index for a Fermion coupled to a $U(1)$ gauge field on $\CP^2$
was derived in \cite{HawkingPope}.

Another bundle that is of interest when $n\ge 2$ 
is the rank $n$ vector bundle $F$ that is inverse to the
generating line bundle $L$ in the sense that $F\oplus L=I^{n+1}$, with $I^{n+1}$ 
the trivial $n+1$ bundle.  Since $F\oplus L$ is  trivial we find that
\begin{eqnarray}
ch(F\oplus L)&=&ch(F)+ch(L)=n+1\\
\Rightarrow ch(F)&=&n+1 -ch(L).
\end{eqnarray}
A second property of the Chern character that will be used below is 
\begin{eqnarray}
ch(F\otimes L)&=&ch(F)\,ch(L)\\
 \Rightarrow ch(F\otimes L^p)&=&(n+1)ch(L^p) - ch(L^{p+1})
\label{eq:ChF}
\end{eqnarray}
The bundle $F$ has structure group $U(n)$ and supports a  
curvature $2$ form of a necessarily topologically non-trivial Yang-Mills gauge 
field on $\CP^n$.

This provides sufficient information to work out the  zero mode structure of the Dirac
operator for spinors transforming under $SU(n)\times U(1)$, in the fundamental
representation of $SU(n)$ and in the background gauge field of $F$.  
More generally we can tensor $F$ with powers of $L$ 
and determine the  zero mode structure  of the Dirac operator for a spinor on $\CP^n$, 
in the fundamental representation of $SU(n)$ and a $U(1)$ charge in the background
field of $F\otimes L^p$. The Dirac index is then denoted by $\nu_{q,{\bf n}}$ where 
\begin{equation}
\nu_{q,{\bf n}}=\int_{\CP^n}ch(F)\,ch(L^p)\,\hat A(\CP^n)=(n+1)\nu_q - \nu_{q-1},
\label{eq:ranknindex}
\end{equation}
where equation (\ref{eq:ChF}) has been used.

Using (\ref{eq:index}) in (\ref{eq:ranknindex})
the index for spinors transforming under the fundamental representation
of $SU(n)$ is
\begin{equation}
\nu_{q,{\bf n}}=(q+n+1)\left(
    \matrix
       {q+n-1 \cr 
       n-1 \cr}
  \right)={(q+n+1)(q+1)\cdots\bigl(q+(n-1)\bigr)\over (n-1)!}.
\label{eq:nindex}
\end{equation}

The presence of $F$ affects the $U(1)$ charge because
\begin{equation}
\int_{S^2}ch(F)\,ch(L^{-q})=\int_{S^2}(n-c_1(L)+\cdots)(1-qc_1(L)+\cdots)=nq+1.
\end{equation}
Since the Chern character involves a trace,
the $U(1)$ generator in the $n\times n$ representation is $(q+{1\over n}){\bf 1}$, where
${\bf 1}$ is the $n\times n$ identity matrix, and the $U(1)$ charge is therefore
$(q+{1\over n})$.  

As different $n$'s are used in the text, when confusion is
possible, we shall distinguish
the $U(1)$ charges for different $\CP^n$'s with a bracketed subscript $n$, as $Y_{(n)}=q+{1\over n}$,
and the different bundles likewise, as $L_{(n)}$ and $F_{(n)}$.

Equations (\ref{eq:index}) and (\ref{eq:nindex}) are the important results
of this analysis.  An immediate consequence is that 
the index vanishes 
for $SU(n)$ singlets 
when $q=-1,\cdots,-n$.  For Fermions in the fundamental representation
of $SU(n)$ the index vanishes for $q=-1,\cdots,-(n-1)$ and $q=-(n+1)$.  
Generically there are no zero
modes for these charges, but even when there are they occur in pairs of opposite
chirality.

\noindent Two further observations, which will be most important in the analysis in the text, 
are 
\begin{itemize}
\item{} Singlets of $SU(n)$ with zero charge have index $\nu_0=1$, while singlets 
with charge $-(n+1)$ have index $\nu_{-(n+1)}=(-1)^n$. 
\item{} In the fundamental representation of $SU(n)$ Fermions coupled to $q=-n$ have 
charge $-n+{1\over n}$ and index $\nu_{-n,{\bf n}}=(-1)^{n+1}$.  
\end{itemize}
\par\vskip0.15\baselineskip

Finally we evaluate the index for spinors in the complex conjugate 
representations---this is relevant to the CPT theorem.
For singlets, since $\overline L=L^{-1}$, 
taking the complex conjugate changes the sign of $p$:  one has $p\mapsto\bar p=-p$, 
and $q\mapsto \bar q =-q-(n+1)$,  giving
\begin{equation}
\nu_{\,\overline q}={1\over n!}(\overline q+1)\cdots(\overline q+n)=(-1)^n \nu_q.
\end{equation}
Note that the $U(1)$ charge is $\overline q =-q-(n+1)$, and not just $-q$, so
$\overline Y_{(n)}=-Y_{(n)}-(n+1)$.

For spinors in the complex conjugate fundamental representation $\overline{\bf n}$
of $SU(n)$ one has,
using $\overline F \oplus \overline L =I^{n+1}$,
%equation (\ref{eq:ChF})  becomes 
\begin{eqnarray}
ch(\overline F)&=&n+1 - e^{-c_1(L)}\\ 
\Rightarrow \qquad\nu_{\,\overline q,\overline{\bf n}}&=&
(n+1)\nu_{\,\overline q} - \nu_{\,\overline q +1}.
\end{eqnarray}
This implies that
\begin{equation}
\nu_{\,\overline q,\overline{\bf n}}
={\overline q(\overline q+2)\cdots(\overline q+n)\over (n-1)!}
=(-1)^n\nu_{q,{\bf n}}.
\label{eq:nbarindex}
\end{equation}
The $U(1)$ charge for $\overline{\bf n}$ is calculated from 
\begin{eqnarray}
\int_{S^2}ch(\overline F)\,ch(L^{-\overline q})&=&
\int_{S^2}(n+c_1(L)+\cdots)(1-{\overline q}c_1(L)+\cdots)\\
&=&n\overline q-1,
\end{eqnarray}
so, dividing by the rank of the matrices as before, the charge is 
$\overline q -{1\over n}=-(q+n+1)-{1\over n}$ so again $\overline Y_{(n)}=-Y_{(n)}-(n+1)$.
To summarise the net effect of complex conjugation is to preserve the
chiralities if $n$ is even and flip them if $n$ is odd and map the charges
to $Y_{(n)}\mapsto\overline Y_{(n)} =-Y_{(n)} -(n+1)$.

\end{document}